\documentclass[conference]{IEEEtran}
\IEEEoverridecommandlockouts
\usepackage{cite}
\usepackage{amsmath,amssymb,amsfonts}
\usepackage{algorithmic}
\usepackage{graphicx}
\usepackage{textcomp}
\usepackage{xcolor}
\usepackage{balance}
\def\BibTeX{{\rm B\kern-.05em{\sc i\kern-.025em b}\kern-.08em
    T\kern-.1667em\lower.7ex\hbox{E}\kern-.125emX}}
\begin{document}

\title{Architecture Matters More Than Scale: A Comparative Study of Retrieval and Memory Augmentation for Financial QA Under SME Compute Constraints\\

}

\author{\IEEEauthorblockN{1\textsuperscript{st} Jianan Liu*}
\IEEEauthorblockA{\textit{Independent Researcher} \\
Austin, TX, USA \\
jiananliu2408@gmail.com*}
\and
\IEEEauthorblockN{2\textsuperscript{nd} Jing Yang}
\IEEEauthorblockA{\textit{Washington University in St. Louis} \\
St. Louis, MO, USA \\
jing.y@wustl.edu}
\and
\IEEEauthorblockN{3\textsuperscript{rd} Xianyou Li}
\IEEEauthorblockA{\textit{New York University} \\
New York, NY, USA \\
xl4230@nyu.edu}
\and
\IEEEauthorblockN{4\textsuperscript{th} Weiran Yan}
\IEEEauthorblockA{\textit{Independent Researcher} \\
Santa Clara, CA, USA \\
yanwr2016@gmail.com}
\and
\IEEEauthorblockN{5\textsuperscript{th} Yichao Wu}
\IEEEauthorblockA{\textit{Northeastern University} \\
Boston, MA, USA \\
wu.yicha@northeastern.edu}
\and
\IEEEauthorblockN{6\textsuperscript{th} Penghao Liang}
\IEEEauthorblockA{\textit{Northeastern University} \\
Boston, MA, USA \\
liang.p@northeastern.edu}
\and
\IEEEauthorblockN{7\textsuperscript{th} Mengwei Yuan}
\IEEEauthorblockA{\textit{Independent Researcher} \\
Milpitas, CA, USA \\
yuanmw1998@gmail.com}
}

\maketitle

\begin{abstract}
The rapid adoption of artificial intelligence (AI) and large language models (LLMs) is transforming financial analytics by enabling natural language interfaces for reporting, decision support, and automated reasoning. However, limited empirical understanding exists regarding how different LLM-based reasoning architectures perform across realistic financial workflows, particularly under the cost, accuracy, and compliance constraints faced by small- and medium-sized enterprises (SMEs) \cite{b1,b2}. SMEs typically operate within severe infrastructure constraints---lacking cloud GPU budgets, dedicated AI teams, and API-scale inference capacity---making architectural efficiency a first-class concern. 

To ensure practical relevance, we introduce an explicit SME-constrained evaluation setting in which all experiments are conducted using a locally hosted 8B-parameter instruction-tuned model without cloud-scale infrastructure. This design isolates the impact of architectural choices within a realistic deployment envelope.

We systematically compare four reasoning architectures—baseline LLM, retrieval-augmented generation (RAG), structured long-term memory, and memory-augmented conversational reasoning—across both FinQA and ConvFinQA benchmarks. Results reveal a consistent architectural inversion: structured memory improves precision in deterministic, operand-explicit tasks, while retrieval-based approaches outperform memory-centric methods in conversational, reference-implicit settings.

Based on these findings, we propose a hybrid deployment framework that dynamically selects reasoning strategies to balance numerical accuracy, auditability, and infrastructure efficiency, providing a practical pathway for financial AI adoption in resource-constrained environments.
\end{abstract}


\section{Introduction}
Financial decision-making increasingly depends on accurate numerical reasoning over structured reports, earnings statements, and tabular disclosures. Financial reporting and financial statement analysis provide standardized representations of corporate financial health, supporting internal managerial analysis, investor evaluation, and regulatory oversight \cite{b3,b4}. Analytical techniques—including ratio analysis, trend analysis, and variance attribution—enable organizations to evaluate performance and guide strategic planning \cite{b5}. In modern enterprise environments, these processes are increasingly augmented by AI-driven tools designed to accelerate reporting cycles and improve analytical accessibility. 

Recent advances in LLMs have demonstrated strong capabilities in natural language understanding and generation \cite{b6,b7}. However, numerical reasoning remains a persistent limitation, particularly in domains requiring arithmetic precision and structured data interpretation \cite{b8,b9}. Financial reasoning poses additional challenges: small computational inaccuracies may produce materially incorrect interpretations, percentage calculations require careful denominator grounding, and structured financial tables often contain multiple semantically similar numeric fields. These weaknesses are amplified in conversational settings, where financial entities, fiscal periods, and reference frames may be implicitly introduced and referenced across multiple dialogue turns. 

To improve factual grounding, retrieval-augmented generation (RAG) architectures have emerged as a promising paradigm for knowledge-intensive tasks \cite{b10,b11}. By dynamically retrieving relevant documents during inference, RAG reduces hallucination and strengthens evidence alignment. In parallel, long-term memory mechanisms have been proposed to enhance contextual continuity and multi-turn reasoning in conversational systems \cite{b12,b13}. While both retrieval and memory architectures aim to address contextual drift and grounding errors, their relative effectiveness in financial numerical reasoning—particularly under practical cost and latency constraints—remains insufficiently explored.

This question carries significant practical importance for small- and medium-sized enterprises (SMEs), which account for approximately 90\% of global businesses and over 50\% of global employment \cite{b1,b2}. Unlike large financial institutions with dedicated AI infrastructure teams, SMEs face concrete deployment constraints: they cannot sustain cloud GPU inference at scale, cannot absorb API-based LLM costs for high-volume financial queries, and must ensure AI systems are deployable on-premise with limited compute budgets. Understanding which architectural choices deliver the strongest accuracy within these constraints is not merely a technical question—it is a prerequisite for responsible AI adoption across the majority of the global economy. To honor this as a first-class experimental concern, we deliberately design our evaluation around a locally hosted 8B-parameter instruction-tuned model without cloud-scale infrastructure. This compute-constrained setup defines the SME feasibility envelope and directly answers what retrieval, memory, and symbolic reasoning architectures can achieve when model size and inference cost are genuinely constrained—as they are for most real-world financial organizations.

In this work, we present a systematic evaluation of financial QA architectures across two complementary benchmarks: FinQA \cite{b9}, which focuses on structured single-turn numerical reasoning over financial reports, and ConvFinQA \cite{b14}, which extends this setting to multi-turn conversational financial analytics requiring contextual entity tracking across dialogue turns. We compare four classes of reasoning systems: (1) baseline LLM reasoning, (2) retrieval-augmented generation (RAG), (3) structured long-term memory reasoning, and (4) memory-augmented conversational reasoning.

Our findings reveal a fundamental contrast between structured numerical reasoning and conversational financial grounding. Structured memory mechanisms improve precision in deterministic reporting environments but degrade under conversational ambiguity, where dynamic retrieval-based re-grounding provides stronger performance and greater robustness. Persistent conversational memory systems often amplify early semantic misalignment and incur higher inference costs without proportional accuracy gains.

We translate these findings into practical deployment guidance through a hybrid routing framework that combines structured symbolic execution for deterministic tasks with retrieval-grounded reasoning for conversational analytics. This framework is particularly suited to SMEs that must balance AI adoption with infrastructure limitations and regulatory transparency requirements.

\section{BACKGROUND \& RELATED WORK}

\subsection{Financial Question Answering Datasets}

\textbf{FinQA.} FinQA \cite{b9} is a benchmark designed to evaluate
numerical reasoning over financial documents. Each example
consists of a financial report excerpt, associated tabular data, and
a question requiring multi-step arithmetic operations. FinQA
provides gold reasoning programs composed of symbolic
operators (e.g., addition, subtraction, division) and operands
grounded in document text or tables, allowing evaluation of both
natural language understanding and program-level reasoning
accuracy.

\textbf{ConvFinQA.} ConvFinQA \cite{b14} extends FinQA into a
conversational setting with multi-turn question-answering
grounded in financial documents. Later questions often reference
prior entities, periods, or computed values implicitly, requiring
conversational state tracking in addition to numerical
computation. Unlike FinQA's isolated queries, ConvFinQA more
closely resembles real-world financial analytics workflows where
analysts iteratively refine queries over reports and tables.

\textbf{Recent Financial QA Benchmarks.} FinanceBench \cite{b15}
evaluates 16 LLM configurations on open-book corporate
disclosure QA, finding that even GPT-4-Turbo with retrieval
incorrectly answered or refused over 80\% of
questions—reinforcing that retrieval grounding in financial
documents remains an unsolved challenge. FinBen \cite{b16}, presented
at NeurIPS 2024, provides a comprehensive multi-task evaluation
across eight financial domains including RAG-based QA, and
finds that model size does not reliably predict task-specific
financial performance—a finding that directly motivates our
architecture-first rather than model-scaling approach.

\subsection{Retrieval-Augmented Generation}

RAG architectures integrate external document retrieval into
the LLM inference process \cite{b10}. By retrieving relevant passages
at runtime, RAG reduces hallucination and strengthens factual
grounding in knowledge-intensive tasks. In financial domains,
retrieval is particularly important because relevant numeric
operands are often embedded in lengthy reports containing
multiple similar figures across periods or entities. However,
retrieval introduces trade-offs related to chunk granularity,
ranking quality, and latency overhead, and does not guarantee
correct operand selection or arithmetic consistency \cite{b11}.
Recent work in financial question answering further highlights that retrieval performance depends critically on balancing robustness and precision, where hybrid document-routed retrieval and reranking strategies can improve grounding quality by adapting retrieval pathways and ranking evidence for financial QA tasks \cite{b17,b18}.

\subsection{Memory-Augmented Large Language Models}

Memory-augmented architectures extend LLMs with
mechanisms for persistent contextual storage across interactions
\cite{b12,b13}. Such systems aim to improve long-range coherence,
reduce repetition, and maintain conversational state across
multi-turn exchanges. In financial analytics contexts, memory
mechanisms may help retain previously referenced entities,
computed values, or fiscal periods. However, persistent memory
may amplify early misinterpretations if incorrect entities or
denominators are stored and reused in subsequent turns.

\subsection{Symbolic and Hybrid Numerical Reasoning}

Symbolic and hybrid neural-symbolic methods address
numerical reasoning tasks requiring explicit arithmetic
computation. Program-of-Thoughts prompting \cite{b19} demonstrates
that externalizing computation into executable programs improves
arithmetic reliability compared to natural language
chain-of-thought reasoning. However, symbolic approaches
typically assume accurate operand selection and explicit
reasoning structure, which conversational settings such as
ConvFinQA may not provide.

\subsection{Positioning Relative to Prior Work.}
Prior studies have evaluated retrieval-augmented generation, 
memory-enhanced dialogue, and symbolic reasoning largely in 
isolation, often focusing on average performance improvements 
within a single task setting. In contrast, our work provides 
a controlled cross-task analysis demonstrating that the 
effectiveness of these paradigms is not absolute but 
conditional on task structure.

We show that architectural performance is governed by the 
interaction between reasoning paradigm and uncertainty type, 
revealing an inversion phenomenon not reported in prior 
literature. This shifts the design question from ``which method 
performs best'' to ``which method aligns with task-specific 
uncertainty,'' providing a more general framework for 
understanding financial QA system behavior.

Building on this perspective, our study introduces three 
distinguishing contributions: 
(1)~\textit{task-architecture alignment analysis}, examining 
how structural task properties interact with retrieval and 
memory mechanisms; 
(2)~\textit{grounding vs.\ persistence trade-offs}, analyzing 
the tension between dynamic retrieval and persistent memory; 
and (3)~\textit{practical deployment considerations}, 
incorporating latency, retrieval overhead, and prompt size 
alongside accuracy.

\section{METHODOLOGY}

\begin{figure*}[t]
\centering
\includegraphics[width=\textwidth]{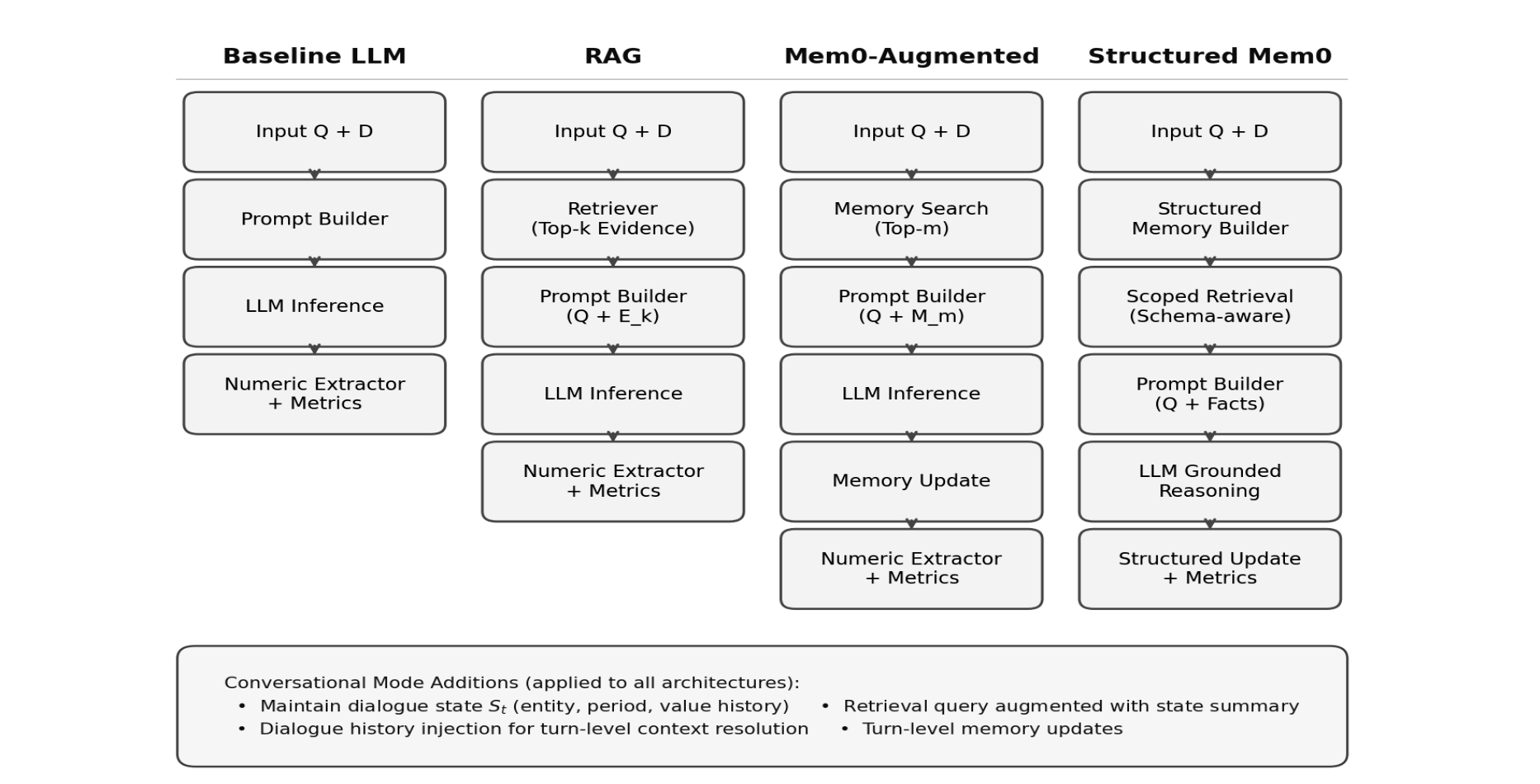}
\caption{Unified architecture overview of the four evaluated systems. All methods share the same input (Q + D) and evaluation layer; they differ in how context is selected, stored, and injected into the prompt. Conversational mode additions (bottom panel) apply across all architectures when evaluating ConvFinQA.}
\label{fig:architecture}
\end{figure*}

This section describes the unified architectural framework
developed for evaluating numerical financial question answering,
illustrated in Fig.~\ref{fig:architecture}. The framework integrates
retrieval-augmented generation, memory-augmented reasoning,
structured state tracking, and symbolic numerical grounding
components. Importantly, the design is task-agnostic and does
not assume a specific dataset structure. Instead, it abstracts
financial numerical QA as a general problem of grounding,
reasoning, and computation over semi-structured financial
documents.

\subsection{Problem Formulation}

We formalize financial numerical question answering as follows.

Let $D$ denote a financial document consisting of textual disclosures and tabular data. Let $Q$ denote a question requiring numerical reasoning grounded in $D$. In conversational settings, a sequence of questions $Q_1, Q_2, \ldots, Q_t$ may be posed over the same document, where later questions depend on prior context.

The objective is to produce a numeric answer $\hat{y} \in \mathbb{R}$ such that:
\begin{equation}
\hat{y} = f(Q, D, S)
\end{equation}

where:
\begin{itemize}
    \item $S$ represents optional conversational state,
    \item $f(\cdot)$ represents a reasoning function implemented via a language model augmented with retrieval and memory mechanisms.
\end{itemize}

Unlike purely extractive QA tasks, numerical financial QA requires:
\begin{enumerate}
    \item \textbf{Operand selection} from tables or text,
    \item \textbf{Operator inference} (e.g., addition, subtraction, ratio, percentage),
    \item \textbf{Multi-step arithmetic execution},
    \item \textbf{Scale normalization} (e.g., thousands, millions),
    \item \textbf{Contextual grounding} across dialogue turns when applicable.
\end{enumerate}

This formulation aligns with prior work on program-based
financial reasoning \cite{b9} and hybrid neural-symbolic numerical QA
\cite{b19}, while extending to conversational and retrieval-augmented
settings.

\subsection{Retrieval and Memory Framework}

Our framework integrates three complementary mechanisms: dynamic retrieval, persistent free-form memory, and structured attribute-value memory.

\subsubsection{Retrieval-Augmented Generation (RAG)}

Retrieval-Augmented Generation \cite{b10} enhances language model inference by incorporating external evidence retrieval. Each financial document is first decomposed into granular facts: pre-text disclosures are split into individual sentences (prefixed PRE:), table rows are serialized as attribute-value pairs (prefixed TABLE:), and post-text lines are similarly factored (prefixed POST:). This structure-aware decomposition is motivated by evidence that granular, element-based chunking substantially improves retrieval precision in financial documents compared to uniform chunking strategies \cite{b20}. Given a query $Q$, a retriever $R$ selects the top-$k$ most relevant facts by cosine similarity over dense embeddings:
\begin{equation}
E = R(Q, D)
\end{equation}

The top-$k$ retrieved facts are concatenated with the question and passed to the language model:
\begin{equation}
\hat{y} = \mathrm{LLM}(Q, E)
\end{equation}

Dynamic retrieval has two important properties: it re-grounds each query independently, and avoids reliance on potentially stale intermediate states. However, retrieval quality depends on fact granularity and ranking accuracy. Missing even a single required operand may cause downstream arithmetic failure.

\subsubsection{Memory-Augmented Reasoning (Mem0-style)}

In addition to retrieval, we implement a persistent memory mechanism inspired by external memory architectures for LLMs \cite{b12}, \cite{b13}. For each example, the full document context is first added to the memory store, then each completed turn's Q/A pair is appended as free-form text. Memory $M_t$ at turn $t$ accumulates these prior interactions:
\begin{equation}
M_t = g(M_{t-1}, Q_{t-1}, \hat{y}_{t-1})
\end{equation}

where $g(\cdot)$ represents the memory update (context addition + Q/A append). At inference time, the most relevant memories for the current query are retrieved and injected alongside the document:
\begin{equation}
\hat{y}_t = \mathrm{LLM}(Q_t, E_t, M_{t-1})
\end{equation}

Persistent memory enables continuity across turns but introduces the possibility of propagating earlier grounding errors if incorrect context is stored. Unlike Structured Mem0, no schema extraction is applied -- content is stored and retrieved as free-form text. Each architecture is evaluated under its default memory scoping behavior: Mem0-Augmented uses a shared memory pool across all dialogs within a run, consistent with its design as a persistent cross-session memory system, while Structured Mem0 scopes memory per dialog via a unique identifier.

\subsubsection{Structured Memory Representation (Structured Mem0)}

Structured memory stores table-derived entities and numeric values as typed attribute-value pairs rather than free-form text. Table rows are serialized using the schema \textit{entity / column = value}. For example:
\begin{itemize}
    \item total volume $|$ 2021 $=$ 637
    \item operating margin $|$ 2020 $=$ 12.5\%
\end{itemize}

Facts are stored directly via embedding without LLM-based extraction (bypassing Mem0's internal inference step), ensuring deterministic, schema-preserving storage. At retrieval time, the top-$k$ facts most similar to the query are selected by cosine similarity, then filtered to remove composite multi-attribute rows in favor of atomic single-attribute facts. This reduces ambiguity compared to free-form textual memory and supports controlled operand reuse, denominator consistency tracking, and explicit entity disambiguation \cite{b9}. In ConvFinQA, memory is scoped per dialog via a unique identifier, preventing cross-dialog fact leakage.

\subsection{Symbolic Numerical Decoding}

To ensure evaluation robustness, we incorporate a lightweight symbolic normalization procedure applied uniformly to the raw outputs of all four architectures during scoring. It is a post-hoc metric computation step only—the LLM always generates free-form text and no architecture receives special treatment. The procedure extracts the last numeric token from the model's free-text output, resolves accounting-style negatives (e.g., parenthetical notation), and applies scale normalization when the question context implies a percentage but the model returns a fraction. Specifically:
\begin{enumerate}
    \item Extract the last numeric token (supporting comma-formatted and parenthetical values),
    \item Normalize fraction-to-percent when question or gold answer implies percent context,
    \item Apply gold-precision-aware tolerance (stricter for multi-decimal answers, more lenient for integer answers).
\end{enumerate}

Because this procedure is applied identically across all methods with no architecture-specific parameters, it isolates architectural effects (retrieval, memory design) from output formatting artifacts rather than conferring any advantage to a particular approach. It produces the ``Corrected'' metric variants reported in Section IV.

\subsection{Conversational Grounding via Dialogue History}

Conversational financial QA introduces implicit references across turns, such as ``what about the following year?'' or ``how does that compare?'' where the referent is not restated. We address this through dialogue history injection: all prior turn Q/A pairs are prepended to the current prompt as explicit context, formatted as:
\begin{itemize}
    \item Turn 1 Q:
    \item Turn 1 A:
    \item $\ldots$ (all previous turns)
\end{itemize}

Given dialogue state $S_{t-1}$ comprising all prior resolved turns, the prompt for turn $t$ becomes:
\begin{equation}
\textit{Prompt}_t = [D, S_{t-1}, Q_t]
\end{equation}

This approach delegates entity tracking and reference resolution to the language model's in-context reasoning, rather than an explicit symbolic resolver. The full prior history is included by default (no fixed window), ensuring that references to values computed in early turns remain accessible throughout the dialogue. Structured Mem0 additionally isolates memory per dialog via a unique identifier to prevent cross-dialog fact leakage.

\subsection{SME-Constrained Experimental Design}

All experiments are deliberately conducted using a locally hosted 8B-parameter instruction-tuned model served via Ollama, without cloud-scale API infrastructure. This is an explicit SME-feasibility design constraint: an 8B-parameter model on commodity hardware represents the realistic upper bound of on-premise deployment for most small- and medium-sized organizations. By fixing this compute ceiling and holding the base model, prompt template, temperature, and decoding parameters constant across all architectural variants, we isolate the effect of retrieval and memory design—and establish which architectural strategies deliver meaningful accuracy gains within the infrastructure envelope available to the organizations that stand to benefit most from financial AI automation.

\subsection{Methodological Summary}

The overall framework combines:
\begin{itemize}
    \item Dynamic fact retrieval (RAG) via cosine similarity over decomposed document facts,
    \item Free-form persistent memory (Mem0-Augmented) accumulating raw context and Q/A history,
    \item Structured attribute-value fact storage with atomic fact filtering (Structured Mem0),
    \item Dialogue history injection for conversational grounding,
    \item Evaluation-time symbolic normalization applied uniformly across all architectures for scale-robust metric computation.
\end{itemize}

These components can be activated independently or combined to produce different architectural variants. This modular design enables systematic comparison of grounding strategies, persistence mechanisms, and memory schemas without changing the core language model.

\subsection{Implementation Details}

All experiments use a locally hosted instance of Llama 3.1 (8B, instruction-tuned) served via Ollama at temperature 0.0. Embeddings are generated using the nomic-embed-text model, also served via Ollama. The RAG pipeline computes cosine similarity directly in Python without an external vector database, caching embeddings to disk for efficiency. Structured Mem0 uses ChromaDB as the vector store backend via the Mem0 library \cite{b13}, with infer=False to bypass LLM-based fact extraction and store facts by direct embedding. Mem0-Augmented uses the same ChromaDB backend but stores and retrieves free-form text.

For both RAG and Structured Mem0, the default retrieval depth is $k = 12$ facts per query. The Structured Mem0 pipeline applies a pattern-based composite-row filter after retrieval: rows exhibiting multi-attribute density are classified as composite and discarded in favor of atomic single-attribute facts, reducing distractor noise in the prompt. The Baseline and RAG ConvFinQA runners include all prior dialogue turns as history context with no fixed window. Structured Mem0 scopes memory per dialog via a unique identifier; Mem0-Augmented operates as a persistent cross-session store, consistent with its default configuration.

\section{FinQA: Single-Turn Numerical Reasoning}

\subsection{Experimental Setup}

FinQA \cite{b9} evaluates multi-step numerical reasoning over financial reports consisting of structured tables and accompanying textual disclosures. Each question is independent, requiring identification of relevant operands and arithmetic operations such as percentage change, ratio computation, or multi-step aggregation. Following the SME-constrained design established in Section III-E, all four architectural variants are evaluated under identical inference conditions on 492 validation examples. No supervised program annotations are used at inference time. Statistical uncertainty for exact match is estimated using Wilson confidence intervals.

We report the following metrics:
\begin{itemize}
    \item \textbf{Exact Match (EM)}: strict numeric equality between prediction and gold.
    
    \item \textbf{Tolerance-based Match (Close)}: prediction within a small relative or absolute tolerance. Both EM and Close are reported in ``Corrected'' variants incorporating the symbolic normalization procedure defined in Section III-C.
    
    \item \textbf{Parse Success Rate}: whether a valid numeric value can be extracted.
    
    \item \textbf{Latency (p50, p95)}: end-to-end inference time in milliseconds.
    
    \item \textbf{Prompt Size}: average input character length (proxy for token cost).
    
    \item \textbf{Judge Correct}: an auxiliary evaluation in which Llama 3.1 8B (instruction-tuned, temperature 0.0, served via Ollama) determines whether the predicted answer matches the gold answer, given a structured comparison prompt that accounts for semantic equivalence and numeric format variations (e.g., 12\% vs. 0.12). The judge model is identical to the inference model to ensure consistency within the SME compute envelope. This metric captures reasoning correctness beyond strict numeric matching and serves as a secondary signal alongside the primary corrected numeric metrics.
\end{itemize}

\subsection{Main Results}

Table~\ref{tab:finqa_results} summarizes performance over 492 validation examples. Several consistent patterns emerge across methods.

\begin{table*}[t]
\caption{FinQA Validation Results (492 samples)}
\label{tab:finqa_results}
\centering
\small
\setlength{\tabcolsep}{6pt}
\begin{tabular}{lcccccc}
\hline
\textbf{Method} & \textbf{Corr. Close} & \textbf{Corr. Exact} & \textbf{Judge Correct} & \textbf{p50 (ms)} & \textbf{p95 (ms)} & \textbf{Avg Prompt Chars} \\
\hline
Baseline LLM & 0.378 & 0.319 & 0.583 & 1466 & 2221 & 4070 \\
RAG & 0.311 & 0.256 & 0.537 & 913 & 1606 & 2165 \\
Mem0-Augmented & 0.293 & 0.236 & 0.514 & 2100 & 4133 & 7063 \\
\textbf{Structured Mem0} & \textbf{0.423} & \textbf{0.354} & 0.565 & 1019 & 4850 & \textbf{1355} \\
\hline
\end{tabular}
\end{table*}

\textbf{Structured Mem0 achieves the highest precision.} Schema-grounded fact storage and constrained retrieval reduce operand ambiguity and scale inconsistencies, yielding the strongest corrected exact and tolerance-based performance. Structured Mem0 achieves the second-lowest median latency (p50 1,019 ms), behind RAG (913 ms), due to compact fact retrieval keeping most prompts small; however, its p95 tail (4,850 ms) is the heaviest of all methods, reflecting occasional long generations on complex multi-step reasoning chains.

\textbf{Baseline LLM performs competitively but is format-sensitive.} Despite no retrieval or memory augmentation, the Baseline achieves the second-highest Corrected Exact (0.319) and the highest Judge score (0.583) among all methods. The 27-point gap between Judge Correct (0.583) and Corrected Exact (0.313) — the largest such gap on FinQA — indicates that a substantial share of baseline answers are semantically correct but fail strict numeric matching due to formatting and scale inconsistencies. This is corroborated by a 48\% format violation rate and 58\% multi-number output rate, suggesting the primary weakness is output formatting rather than underlying arithmetic reasoning.

\textbf{RAG improves retrieval efficiency but not deterministic arithmetic precision.} 
RAG achieves the lowest p50 latency of all methods (913 ms) with the smallest prompt footprint among retrieval methods. 
However, both its Corrected Exact (0.256) and Corrected Close (0.311) fall below Baseline LLM (0.319 / 0.386) and Structured Mem0 (0.354 / 0.427), 
making RAG the weakest architecture on FinQA despite its latency advantage. 
The narrow gap between RAG's Corrected Exact (0.256) and Corrected Close (0.311) — only 0.055 — indicates that RAG errors are not primarily scale or formatting artifacts 
but reflect genuine operand selection failures: the retrieved context leads the model to the wrong numeric value entirely, 
not merely to a mis-scaled version of the right one. 
This is consistent with distractor rows introduced by near-duplicate table entries retrieved at $k = 12$ without composite-row filtering, 
and is most pronounced on percentage-answer questions where denominator disambiguation is required (see Fig.~\ref{fig:appendix_finqa_type} in the Appendix) \cite{b18}. 
Notably, RAG ranks second on Judge Correct (0.537) despite its low exact score, a mild instance of the fluency-accuracy divergence observed more sharply in ConvFinQA: 
retrieved context produces coherent, well-formed responses that nonetheless anchor to the wrong numeric operand. 
These findings suggest that retrieval precision—not retrieval coverage—is the binding constraint for RAG on deterministic FinQA tasks.

\textbf{Mem0-Augmented increases contextual bandwidth but degrades accuracy.} Unstructured memory injection substantially increases prompt size ($\sim$7,063 characters average) and latency without improving corrected performance, highlighting scalability limitations of naive context expansion.

\subsection{Why Structured Memory Helps in FinQA}

The improvements from Structured Mem0 arise from task-architecture alignment. First, FinQA questions reference clearly identifiable financial metrics and periods, and structured memory stores facts as typed entity-period-metric tuples, reducing confusion when multiple similar values appear within the same document. Second, constraining retrieval to schema-validated facts reduces operand noise while symbolic normalization stabilizes scale interpretation. Third, because FinQA is single-turn, persistent memory does not introduce cross-turn error propagation—it acts as a curated knowledge substrate rather than a dynamic dialogue state.

\subsection{Model-Bounded Performance Ceiling}

Despite architectural differences, all methods remain below 50\% corrected exact accuracy. To contextualize this: supervised program induction methods evaluated on larger models in the original FinQA work achieve substantially higher absolute scores— \cite{b9} report $\sim$68\% execution accuracy using gold program supervision with a fine-tuned retriever. Our results are not intended to compete with that setting. This study makes no state-of-the-art claim; its contribution is architectural comparison within a fixed, SME-feasible compute envelope. The relevant question here is not ``how close to supervised state-of-the-art (SOTA) can an 8B model get?'' but rather ``given a fixed 8B locally-hosted model, which architecture delivers the strongest accuracy?''—a question that prior work does not answer. Within this range, architectural choice accounts for meaningful variation: the gap between the weakest and strongest architecture on FinQA corrected exact match is 11.8 percentage points—achieved with no change in model, hardware, or inference cost. This demonstrates that for organizations operating under realistic compute constraints, architecture selection is a more accessible and impactful lever than model scaling.

\section{ConvFinQA: Conversational Financial QA}

\subsection{Task Differences vs. FinQA}

ConvFinQA extends FinQA by introducing multi-turn dialogue context, implicit metric references (``what about the following year?''), denominator ambiguity in percentage questions, and cross-turn dependency on prior grounding decisions. Unlike FinQA, ConvFinQA does not explicitly restate operands in each turn. The model must infer which metric is under discussion, which year or row is referenced, and whether the question concerns an absolute value, difference, or percentage change. This introduces entity identity uncertainty rather than purely numeric uncertainty.

\subsection{Experimental Setup}

We evaluate 500 ConvFinQA validation dialogs comprising 1,490 total turns. All methods use the same locally hosted Ollama 8B backend established in Section III-E.

\subsection{Main Results}

We report the following metrics: \textbf{Exact Match} (strict numeric equality); \textbf{Auto Close} (tolerance-based match with auto-fix formatting correction, primary metric); \textbf{95\% CI} (Wilson confidence interval on Auto Close); \textbf{Judge} (secondary semantic correctness via LLM adjudication); \textbf{p50 latency} (median inference time); and \textbf{Avg Tokens} (average prompt token count, proxy for inference cost).

Table~\ref{tab:convfinqa_results} summarizes turn-level performance over 500 validation dialogs (1,490 total turns), highlighting consistent differences across architectures under conversational grounding.

\begin{table*}[t]
\caption{ConvFinQA Turn-Level Results (500 dialogs, 1,490 turns)}
\label{tab:convfinqa_results}
\centering
\small
\setlength{\tabcolsep}{6pt}
\begin{tabular}{lcccccc}
\hline
\textbf{Method} & \textbf{Exact Match} & \textbf{Auto Close} & \textbf{95\% CI} & \textbf{Judge} & \textbf{p50 (ms)} & \textbf{Avg Tokens} \\
\hline
\textbf{RAG} & \textbf{43.49\%} & \textbf{52.75\%} & [50.2\%, 55.3\%] & 57.52\% & 2517 & \textbf{717} \\
Baseline LLM & 41.95\% & 48.46\% & [45.9\%, 51.0\%] & 52.08\% & 1317 & 1044 \\
Structured Mem0 & 38.19\% & 46.64\% & [44.1\%, 49.2\%] & 48.86\% & 2625 & 705 \\
Mem0-Augmented & 36.17\% & 43.22\% & [40.7\%, 45.8\%] & 63.76\% & 2951 & 2088 \\
\hline
\end{tabular}
\end{table*}

\begin{figure}[t]
\centering
\includegraphics[width=\columnwidth]{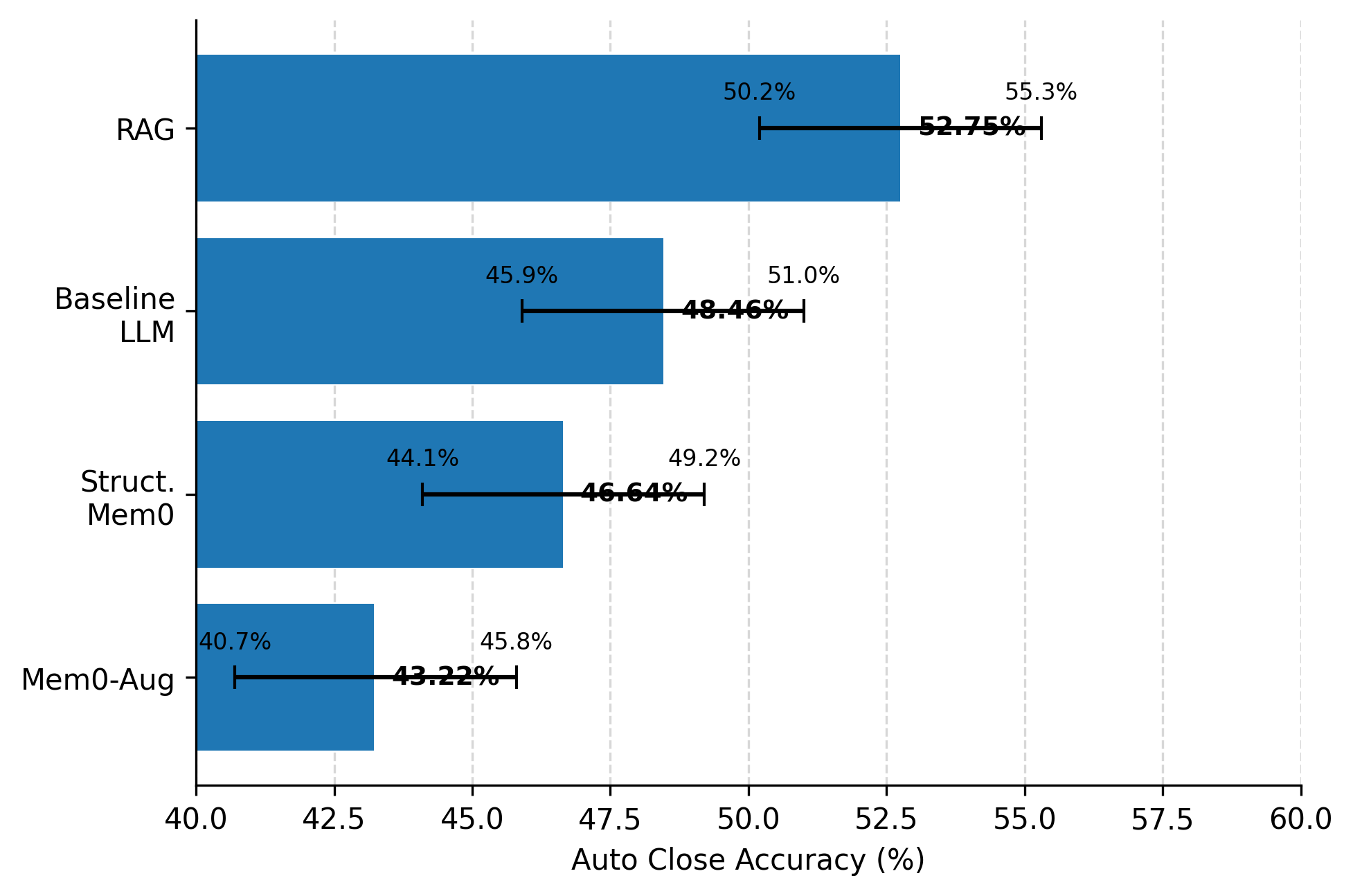}
\caption{ConvFinQA close accuracy with 95\% Wilson confidence intervals. RAG achieves the highest close accuracy, and its confidence interval is separated from those of the other architectures, supporting the statistical reliability of the observed ranking.}
\label{fig:convfinqa_ci}
\end{figure}

RAG dominates numerically, achieving +4.3\% absolute over Baseline and +9.5\% over Mem0-Augmented in close accuracy. Structured memory underperforms even baseline. Mem0-Augmented performs worst overall at 43.22\% close accuracy while consuming nearly 3x the tokens of RAG, exhibiting the heaviest large-error tail due to semantic mis-grounding. Fig.~\ref{fig:convfinqa_ci} plots these results with 95\% Wilson confidence intervals, confirming that the RAG advantage is statistically reliable at this sample size.

\subsection{Why Structure and Memory Hurt in ConvFinQA}

Unlike FinQA, ConvFinQA penalizes early commitment. Structured Mem0 normalizes a metric, canonicalizes year identity, and persists it across turns. If this first commitment is incorrect, all subsequent turns remain internally consistent arithmetically but anchor to the wrong entity identity, leading to monotonic degradation. In contrast, RAG re-retrieves each turn, allowing accidental recovery and benefiting from redundancy in table exposure.

\begin{figure}[t]
\centering
\includegraphics[width=\columnwidth]{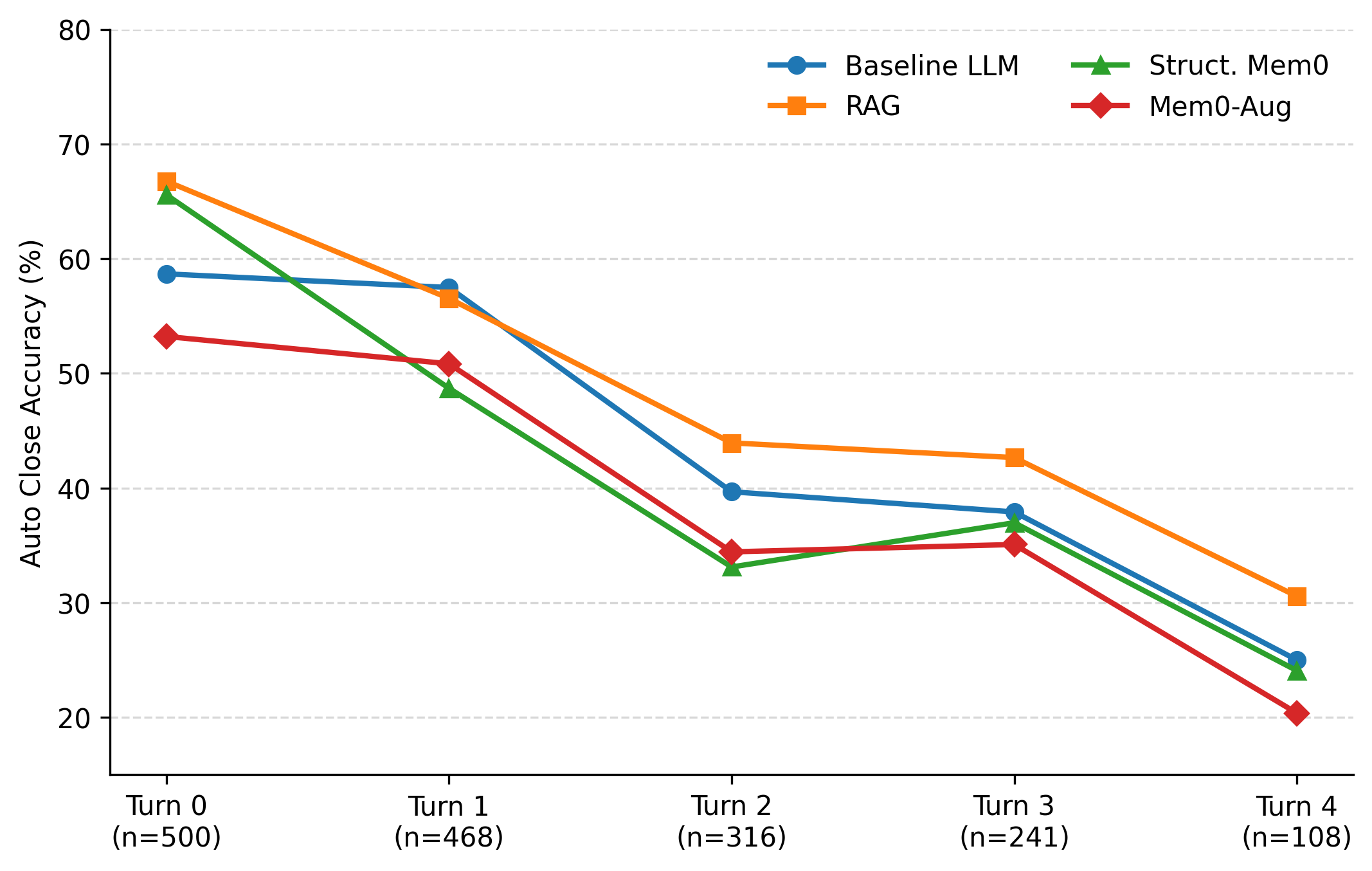}
\caption{ConvFinQA close accuracy by conversation turn index. Performance declines with turn depth across all methods, with RAG maintaining the strongest robustness under multi-turn conversational grounding.}
\label{fig:turn_index_accuracy}
\end{figure}

Table~\ref{tab:turn_index_accuracy} shows accuracy by turn index across all methods, with the same trend visualized in Fig.~\ref{fig:turn_index_accuracy}.

\begin{table}[t]
\caption{Accuracy by Turn Index (ConvFinQA)}
\label{tab:turn_index_accuracy}
\centering
\small
\setlength{\tabcolsep}{5pt}
\begin{tabular}{ccccc}
\hline
\textbf{Turn} & \textbf{Baseline} & \textbf{RAG} & \textbf{Struct. Mem0} & \textbf{Mem0-Aug} \\
\hline
0 & 58.67\% & 66.75\% & 65.56\% & 53.21\% \\
1 & 57.48\% & 56.53\% & 48.69\% & 50.83\% \\
2 & 39.67\% & 43.93\% & 33.11\% & 34.43\% \\
3 & 37.91\% & 42.65\% & 36.97\% & 35.07\% \\
4 & 25.00\% & 30.56\% & 24.07\% & 20.37\% \\
\hline
\end{tabular}
\end{table}

RAG degrades gradually across turns, while memory-based approaches degrade sharply after Turn 1, with Mem0-Augmented showing the steepest collapse. Table~\ref{tab:cascade_failure_rates} shows cascade failure rates: the median cascade rate of 100\% across all methods indicates that once the first turn fails, later turns almost never recover. The mean cascade rate reveals a ranking (RAG 68.99\% $<$ Structured Mem0 70.14\% $<$ Baseline 73.84\% $<$ Mem0-Aug 74.91\%), confirming that retrieval reduces error propagation while structured persistence amplifies early mis-grounding.

\begin{table}[t]
\caption{Cascade Failure Rates (ConvFinQA)}
\label{tab:cascade_failure_rates}
\centering
\small
\setlength{\tabcolsep}{6pt}
\begin{tabular}{lcc}
\hline
\textbf{Method} & \textbf{Mean Cascade} & \textbf{Median Cascade} \\
\hline
\textbf{RAG} & \textbf{68.99\%} & \textbf{100\%} \\
Structured Mem0 & 70.14\% & 100\% \\
Baseline LLM & 73.84\% & 100\% \\
Mem0-Augmented & 74.91\% & 100\% \\
\hline
\end{tabular}
\end{table}

\subsection{Fluency-Accuracy Divergence}

\begin{figure*}[t]
\centering
\includegraphics[width=\textwidth]{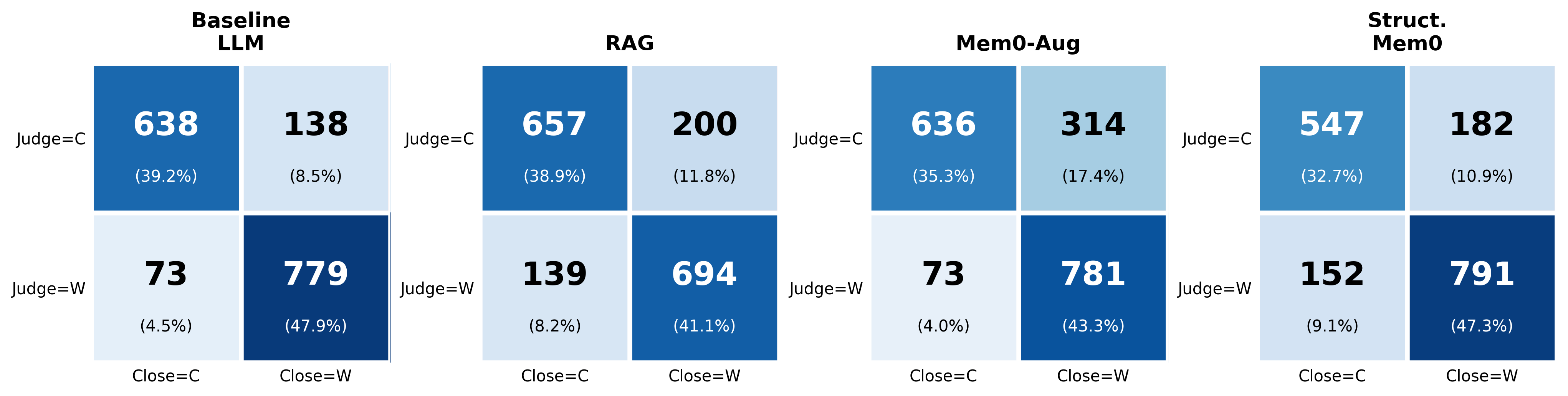}
\caption{Judge versus close confusion matrices across architectures in ConvFinQA. The top-right cell corresponds to fluency--accuracy divergence (Judge = Correct, Close = Wrong). Mem0-Augmented exhibits the largest divergence, indicating that memory-heavy architectures produce fluent but numerically incorrect responses more frequently. Counts and percentages are shown per cell.}
\label{fig:fluency_divergence}
\end{figure*}

As shown in Table~\ref{tab:convfinqa_results}, Mem0-Augmented achieves the highest Judge score (63.76\%) despite the lowest numeric-close accuracy (43.22\%) -- a gap of over 20 percentage points. This is the sharpest divergence observed across all methods and both benchmarks. Analysis reveals 314 turns where the judge scored ``correct'' but numeric-close scored ``incorrect'' for Mem0-Augmented alone—2.3x the rate of Baseline LLM (138 turns); the full per-method breakdown is shown in Fig.~\ref{fig:fluency_divergence}. Memory-heavy architectures produce more coherent, plausible-sounding responses with fewer formatting violations and higher semantic fluency, yet they simultaneously anchor to the wrong metric or denominator. This fluency-accuracy divergence is a critical warning for practitioners: Judge-style evaluations alone will systematically overestimate the real-world reliability of persistent memory architectures in financial QA. Numeric grounding metrics must be reported alongside fluency scores.

\subsection{Latency and Token Economics}

Table~\ref{tab:efficiency_metrics} presents efficiency metrics. Retrieval cost is negligible relative to generation cost. RAG achieves the strongest cost-accuracy balance: despite higher latency than baseline, it consumes fewer tokens per turn (717 avg) while achieving the highest numeric accuracy. Mem0-Augmented consumes $\sim$3x the tokens of RAG (2,088 vs. 717) while delivering the lowest numeric accuracy. Baseline LLM has the lowest latency (1,317 ms p50) but the highest token count among non-memory methods.

\begin{table}[t]
\caption{Efficiency Metrics (ConvFinQA)}
\label{tab:efficiency_metrics}
\centering
\small
\setlength{\tabcolsep}{6pt}
\begin{tabular}{lccc}
\hline
\textbf{Method} & \textbf{p50 (ms)} & \textbf{p95 (ms)} & \textbf{Avg Tokens} \\
\hline
Baseline LLM & 1317 & 2229 & 1044 \\
RAG & 2517 & 5123 & 717 \\
Structured Mem0 & 2625 & 5226 & 705 \\
Mem0-Augmented & 2951 & 4221 & 2088 \\
\hline
\end{tabular}
\end{table}

\section{Cross-Dataset Synthesis}

Sections IV and V reveal a systematic architectural inversion between FinQA and ConvFinQA. This inversion reflects a fundamental interaction between task ambiguity structure and architectural inductive bias, which we characterize as follows.

In FinQA, uncertainty is primarily computational: operands, metric identities, and temporal references are explicitly specified within a single turn. Under these conditions, structured memory and symbolic normalization improve precision by stabilizing arithmetic execution.

In ConvFinQA, uncertainty shifts to referential grounding. Questions frequently omit explicit metric identity and require cross-turn entity resolution, introducing ambiguity in denominator selection and temporal context. In this setting, persistent representations can amplify early misalignment, while retrieval-based re-grounding improves recoverability.

This contrast demonstrates a broader principle: \emph{architectural persistence is beneficial when semantic identity is stable, but harmful when semantic identity is ambiguous}. Across both datasets, arithmetic reasoning remains comparatively robust. The primary performance ceiling arises not from numeric execution but from entity alignment under ambiguity. The $\sim$50--55\% accuracy plateau observed across architectures on ConvFinQA reflects the difficulty of unsupervised conversational grounding rather than insufficient computational capacity. This inversion is stable across exact and tolerance-based evaluation metrics, turn-depth analyses, cascade failure diagnostics, and efficiency measurements—indicating task-architecture interactions rather than metric artifacts.

\subsection{Mechanistic Interpretation via Uncertainty Decomposition}

\textbf{Architecture-Specific Behavior.}
The key difference between architectures lies in when they commit to interpreting entities and numeric references. Retrieval-augmented generation (RAG) defers commitment: each query retrieves evidence independently, selecting relevant facts at inference time and discarding prior context. This behavior introduces variability in FinQA, where distractor rows may lead to incorrect operand selection, but becomes advantageous in ConvFinQA, where incorrect intermediate reasoning can be corrected through re-grounding in subsequent turns.

Structured Mem0 adopts the opposite strategy by constructing a normalized representation of entities and values upfront. In deterministic settings such as FinQA, this enables precise and stable operand selection. However, in conversational settings, implicit references (e.g., “the following year”) require reinterpretation of context. Early commitment leads to persistent misalignment, as incorrect entity bindings are reused across turns.

Mem0-Augmented further introduces ambiguity by storing raw question-answer pairs. Retrieval from such memory is driven by semantic similarity rather than entity-level alignment, causing numerically distinct but linguistically similar entries to interfere with each other.

\textbf{Uncertainty Decomposition.}
We interpret these behaviors through a decomposition of task uncertainty into two components: (i) \emph{computational uncertainty}, arising from arithmetic execution, and (ii) \emph{referential uncertainty}, arising from ambiguity in entity and context grounding.

In FinQA, referential uncertainty is minimal and computational uncertainty dominates, favoring architectures that enforce structured constraints and persistent representations. In ConvFinQA, referential uncertainty dominates, and architectures that allow dynamic re-grounding provide greater robustness.

This unified perspective explains why architectural choice—rather than model scale—governs performance: different architectures implicitly bias toward reducing different sources of uncertainty.

\section{Industry \& System Design Implications}

\subsection{Financial QA as Dual-Mode Infrastructure}

Modern financial workflows consist of two structurally distinct reasoning environments: deterministic reporting pipelines (variance analysis, reconciliation, audit calculations, compliance checks) and conversational analytic interactions (executive Q\&A, exploratory analysis, performance interpretation). Our evaluation demonstrates that these modes require distinct architectural biases—structured symbolic execution for operand-explicit tasks and retrieval-driven re-grounding for reference-implicit tasks. Financial AI systems should therefore be deployed as dual-mode analytic infrastructure with dynamic routing across reasoning paradigms rather than as monolithic conversational agents.

\subsection{Cost-Accuracy Trade-offs in Production Systems}

Persistent memory architectures increased token consumption by up to $\sim$3x without improving conversational accuracy, while retrieval-grounded systems achieved the strongest cost-to-accuracy balance with negligible retrieval latency relative to LLM inference cost. For organizations processing thousands of analytic queries per day, even modest token reductions (25--40\%) translate into substantial annual cost savings at API-based inference scale. For SMEs, where IT budgets are limited and finance teams are small, cost-efficiency is critical—retrieval-first architectures provide a practical path toward AI adoption without enterprise-scale infrastructure investment.

\subsection{Governance, Compliance, and Auditability}

Financial AI systems must satisfy regulatory requirements related to transparency, traceability, and reproducibility. Retrieval-grounded architectures inherently link outputs to source financial documents, improving auditability in alignment with emerging regulatory expectations for explainable AI in financial services. Symbolic numeric post-processing further strengthens compliance by enabling deterministic verification independent of generative reasoning. Conversational memory architectures, by contrast, risk compounding early mis-grounding and may obscure evidence lineage, complicating audit reconstruction.

\subsection{Task-Aware Architecture Selection: Empirical Validation}

The architectural inversion identified in Sections IV--VI is not merely descriptive—it is actionable. We validate this directly by simulating a task-aware router: a system that routes deterministic single-turn financial queries to Structured Mem0 and conversational multi-turn queries to RAG, using task structure (benchmark membership) as the routing signal.

\textbf{Routing Decision Formalization.}
We formalize architecture selection as a routing function $f(q, h)$, where $q$ is the current query and $h$ denotes the dialogue history:
\[
f(q, h) =
\begin{cases}
\text{Structured Mem0}, & \text{if } h = \emptyset \\
\text{RAG}, & \text{if } h \neq \emptyset
\end{cases}
\]

This formulation captures the primary structural signal identified in our experiments: the presence of dialogue history serves as a reliable proxy for referential ambiguity.

\textbf{Practical Routing Rule.}
The routing rule can be stated precisely in operational terms.

If there is no prior conversation history and the current question explicitly specifies the metric and time period, Structured Mem0 is used. If there is any prior conversation history, or if the question depends on context from earlier turns, RAG is selected.

The most reliable signal is whether a dialogue history exists. This criterion is computationally trivial and effectively separates single-turn deterministic queries from multi-turn conversational queries without requiring semantic parsing of the question.

A secondary signal is whether the question is self-contained. For example, “what was the operating margin in 2021?” contains all necessary information for deterministic retrieval, whereas “how does that compare to last year?” depends on prior context. In self-contained single-turn queries, Structured Mem0 performs best; in all other cases, RAG provides more robust performance.

Table~\ref{tab:routing_experiment} reports combined accuracy under four routing conditions: each single architecture applied uniformly across both benchmarks, a keyword-heuristic router, and an oracle router. The oracle router selects the empirically strongest architecture per task type (Structured Mem0 for FinQA, RAG for ConvFinQA) and represents the upper bound achievable with perfect task-type classification.

\begin{table*}[t]
\caption{Routing Experiment --- Combined Accuracy Across FinQA + ConvFinQA ($n$=1,982)}
\label{tab:routing_experiment}
\centering
\small
\setlength{\tabcolsep}{6pt}
\renewcommand{\arraystretch}{1.1}
\begin{tabular}{lcccc}
\hline
\textbf{Method} & \textbf{FinQA Close} & \textbf{Conv. Close} & \textbf{Combined Close} & \textbf{vs. Best Single} \\
\hline
Baseline LLM & 0.386 & 0.477 & 0.455 & --- \\
RAG & 0.311 & 0.534 & 0.479 & +0.0pp \\
Mem0-Augmented & 0.309 & 0.476 & 0.434 & -4.5pp \\
Structured Mem0 & 0.427 & 0.469 & 0.458 & +0.3pp \\
Keyword Heuristic (RAG + Struct. Mem0) & 0.388 & 0.507 & 0.478 & +0.0pp \\
\hline
\textbf{Oracle Router (Struct.\ Mem0 $\rightarrow$ FinQA; RAG $\rightarrow$ ConvFinQA)} & \textbf{0.427} & \textbf{0.534} & \textbf{0.508} & \textbf{+2.9pp} \\
\hline
\end{tabular}
\end{table*}

The oracle router achieves a combined close accuracy of 50.8\%—a +2.9 percentage point improvement over the best single architecture (RAG, 47.9\%), with no change in model or inference cost. This confirms that architectural inversion is not a theoretical observation but a practically exploitable property: selecting the right architecture for the task type yields consistent gains across both benchmarks simultaneously.

The keyword-heuristic router (routing questions containing implicit reference language to RAG, explicit operand questions to Structured Mem0) does not improve over RAG alone. This is informative: single-turn FinQA questions also contain words such as ``change,'' ``ratio,'' and ``percentage,'' making lexical features insufficient for distinguishing task structure. Effective routing therefore requires structural signals—specifically, whether a dialogue history is present—rather than question-level keyword matching. This aligns directly with the routing function defined above, where dialogue history serves as the primary decision signal. This finding directly motivates the lightweight task-structure classifier proposed in Section VIII-A as a practical path to realizing oracle-level routing gains at deployment.

\subsection{Practical Deployment Roadmap}

Based on cross-dataset findings, we propose a four-phase staged deployment roadmap as design guidance. \textbf{Phase 1 (Retrieval-First Baseline):} deploy evidence-grounded QA with optimized row-level retrieval and entity filtering, avoiding persistent conversational memory initially. \textbf{Phase 2 (Deterministic Numeric Verification):} add symbolic normalization, percent/fraction harmonization, and post-generation arithmetic checking. \textbf{Phase 3 (Context-Aware Routing):} dynamically route deterministic tasks to structured symbolic execution and ambiguous conversational queries to retrieval-based grounding, using dialogue-history presence as the primary routing signal. \textbf{Phase 4 (Supervised Enhancement, Optional):} introduce supervised pointer grounding only when conversational query complexity justifies the cost.

\subsection{Breaking the 50\% Ceiling}

Across ConvFinQA, accuracy stabilizes near $\sim$50--55\% under unsupervised architectures. This ceiling reflects conversational entity ambiguity rather than arithmetic weakness, suggesting that breaking it likely requires supervised entity-grounding models, explicit pointer-resolution training, and multi-hypothesis conversational state tracking. Scaling model size alone is unlikely to resolve denominator ambiguity or referential drift—supervised entity identification models outperform LLM-based few-shot approaches on entity tracking tasks even when the underlying LLM is large \cite{b21}, suggesting that architectural inductive biases for entity persistence require targeted supervision beyond generic pretraining. For organizations evaluating AI investments, this implies that performance improvements beyond moderate accuracy require labeled data and task-specific training, not merely larger foundation models.

\section{Conclusion and Future Work}

This work presented a systematic evaluation of four financial QA architectures across FinQA and ConvFinQA, revealing a fundamental architectural inversion. In deterministic, operand-explicit environments, structured symbolic reasoning and memory-enhanced execution improve numeric precision. In conversational, reference-implicit environments, retrieval-driven re-grounding consistently outperforms persistent memory systems. The same architectural persistence that improves precision under semantic stability degrades robustness under referential ambiguity.

Across both datasets, arithmetic computation was not the primary bottleneck. Performance ceilings emerged from conversational grounding challenges—denominator disambiguation, entity drift, and early semantic commitment. These results shift the design perspective from model scaling to architectural alignment: financial QA performance depends less on foundation model size and more on whether architectural bias matches the task's uncertainty structure.

Beyond benchmark comparison, this study contributes a controlled cross-dataset contrast revealing task-dependent architectural inversion, an uncertainty-alignment framing explaining when persistence helps and when it harms, cost-aware evaluation incorporating latency and token economics, and a deployment-oriented design framework applicable to enterprises and SMEs. All code, configurations, and per-sample result logs are released to support reproducibility (see Section III-G).

\subsection{Limitations}

\textbf{Single-Model Scope.} All experiments use a single base model: Llama 3.1 8B instruction-tuned, served locally via Ollama. While this reflects a deliberate SME-feasibility design constraint, it raises the question of whether the architectural inversion—Structured Mem0 winning on FinQA, RAG winning on ConvFinQA—generalizes across other models in the 7--13B parameter range (e.g., Mistral 7B, Qwen2 7B, Gemma 2 9B). We cannot confirm from the current data that the inversion is model-agnostic; it is possible that models with stronger in-context entity tracking or longer effective context windows would narrow the ConvFinQA gap for memory-based architectures. This is an explicit limitation rather than an implicit one: the contribution of this study is architecture comparison within a fixed, SME-relevant compute envelope, and extending the analysis to additional base models is a direct and tractable direction for future work.

\textbf{Benchmark Coverage.} Findings are based on two benchmarks—FinQA and ConvFinQA—that are grounded in public financial filings and share overlapping document structure. Generalization to other financial QA domains (e.g., earnings call transcripts, regulatory filings, internal management accounts) may differ, particularly where document structure, table density, or implicit reference patterns diverge from the benchmark distributions.

\textbf{Generalization Beyond Model Scale.}
Using a local 8B model was a deliberate choice to reflect what most SMEs can realistically deploy; however, it is important to consider whether the observed findings generalize to larger or cloud-hosted models.

We argue that the architectural patterns identified in this study are structural rather than parametric. The observed inversion—Structured Mem0 outperforming in FinQA and RAG outperforming in ConvFinQA—arises from whether early commitment to an entity interpretation is safe or error-prone. This is fundamentally determined by task structure, not model size. While larger models may improve absolute accuracy, they are unlikely to eliminate the directional effects observed: persistent memory may still propagate early misalignment, while retrieval-based re-grounding enables recovery in conversational settings.

Furthermore, scaling model size may amplify certain failure modes. Larger models produce more fluent and coherent outputs, which can increase the gap between perceived correctness and actual numeric accuracy. As a result, the divergence between judge-based evaluation and true numeric correctness may become more pronounced at scale, reinforcing the importance of robust evaluation metrics.

From a systems perspective, the cost-efficiency trade-offs also generalize. Mem0-Augmented consumes significantly more tokens than retrieval-based approaches while delivering lower accuracy. Since token costs scale with model size in API-based deployments, the economic advantage of retrieval-first architectures is likely to increase rather than diminish in larger-scale systems.

Taken together, these observations suggest that the 8B constraint defines a lower bound on deployment feasibility rather than an upper bound on applicability. While absolute performance levels may vary across models, the relative architectural behaviors and design implications are expected to persist across scales.

\subsection{Future Work}

\textbf{Supervised Conversational Grounding.} The conversational accuracy ceiling in ConvFinQA suggests targeted supervision—particularly pointer-grounding and denominator disambiguation training tailored to financial discourse—may substantially improve robustness without requiring full program induction.

\textbf{Adaptive Hybrid Routing Architectures.} The oracle routing experiment in Section VII-D demonstrates a +2.9pp combined accuracy gain from task-aware architecture selection, with no model or hardware changes. Realizing this gain in deployment requires a lightweight task-structure classifier—distinguishing single-turn deterministic queries from multi-turn conversational ones based on dialogue-history presence and context window signals. Developing and evaluating such a classifier represents a direct and tractable extension of this work.

\textbf{Large-Scale Operational Validation.} While benchmark datasets provide controlled evaluation, validating hybrid architectures within live financial reporting and decision-support environments would further quantify cost-accuracy trade-offs, auditability benefits, and productivity gains at deployment scale.

\bibliographystyle{IEEEtran}
\bibliography{references}

\balance
\suppressfloats[t]

\par\vspace{4pt}
\begin{center}
\normalfont\small
APPENDIX\\
Supplementary Figures
\end{center}
\vspace{2pt}

\setcounter{figure}{0}
\renewcommand{\thefigure}{A\arabic{figure}}

\noindent
\refstepcounter{figure}
\begin{minipage}{\columnwidth}
\centering
\includegraphics[width=0.95\columnwidth,trim={0 180 0 120},clip]{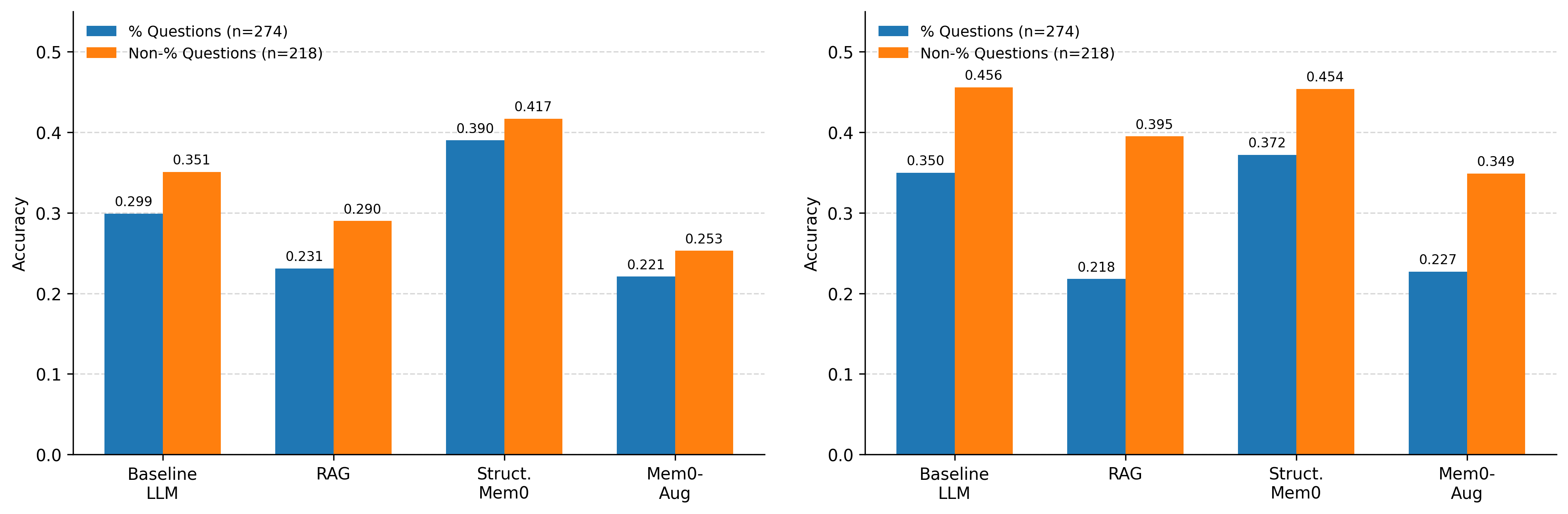}
\vspace{4pt}

{\small \textbf{Fig.~\thefigure.} FinQA corrected accuracy by question type. Performance is reported separately for percentage-based and non-percentage questions across all methods. Structured Mem0 achieves the highest accuracy across both categories, while RAG underperforms due to operand selection errors, particularly in percentage questions requiring denominator disambiguation.}
\label{fig:appendix_finqa_type}
\end{minipage}

\end{document}